# A Comprehensive Optimization Method for Commercial Building Loads with Renewable Generation and Energy Storage from Utility Rate Structure Perspective

A S M Jahid Hasan, *Student Member, IEEE,* Jubair Yusuf, *Student Member, IEEE,* Sadrul Ula, *Senior Member, IEEE*
Department of Electrical and Computer Engineering
University of California Riverside
CA, USA
ahasa006@ucr.edu, jyusu001@ucr.edu, sula@cert.ucr.edu

*Abstract*—To accommodate the changes in the nature and pattern of electricity consumption with the available resources, utility companies have introduced a variety of rate structures over the years. This paper develops a comprehensive optimization method that addresses the diversity of utility rate structures of commercial buildings. It includes a general set of constraints that can be used for any system with a building load, a renewable source, and a battery energy storage system (BESS). A cost function is formulated for each type of rate structure that can be exercised by a utility on a commercial building. A novel algorithm is developed to apply the optimization model and generate the desired optimal outputs by using the appropriate cost function. The results for several building loads and rate structure types were obtained and compared. A sensitivity analysis was done on the optimization model based on the changes in the rates using historical data. The results exhibit that adding BESS is more effective for buildings with lower load factor and CPP rate structures in comparison to the buildings with flat energy rates. Savings from adding renewables such as solar is primarily influenced by energy charges whereas additional benefits from BESS are dominated by demand charges. These results can help a customer with deciding on the different rate structure options and resource planning of their renewable generation and energy storage. Utilities may also benefit from this work by designing a unified rate structure, considering the increasing renewable penetration and BESS deployment in the grid.

## I. INTRODUCTION

### A. Motivation

The nature and pattern of electrical loads have changed significantly over the years. The recent adoption of high renewable penetration has made the change even more severe and rapid. The utilities have developed numerous types of electrical rate structures accordingly that would facilitate their operation with the existing generation, transmission, and distribution resources. Concepts such as Time of Use (TOU) and different types of energy and demand charges have been introduced to remedy the limitations of these resources. Additionally, highly distinctive load patterns and consumption amount have urged the utilities to include separate rate structures for residential and industrial/commercial sectors. Commercial buildings being the one of the largest electricity users in the U.S. and consuming about 35% of the total electricity consumption [1]; the users are immediately affected from any change in the electrical rate structure. As one of the largest electricity users in the U.S., building loads offer great potential for optimization with the help of renewables and energy storage that would largely benefit both the customers behind the meter and utility operators. But these extremely diverse and highly complex rate structures as well as the pattern of the loads with building size and types create a major challenge for formulating an appropriate optimization strategy. On top of that, the price of the energy and demand charges are continuously changing every year, making it even more challenging. Therefore, in order to design an optimal strategy for reducing building loads with renewable generation and energy storage, the rate structure is an important component that needs to be considered.

In addition to the increment of renewable consumption, a unified rate structure can reduce the uncertainties associated with various types of charges for the customers behind the meter. The customers often cannot decide the best rate structure option for themselves with a very little knowledge of how these complicated charges are applied to different energy consumption behaviors. This lack of knowledge leads to the yearly price increment made by the utilities as the prosumers cannot make the best use of their resources to provide an optimal load management. Hence a comprehensive utility rate structure can be beneficial for both customers and utilities when optimality is ensured for the overall system. A thorough analysis of optimization with renewable and energy storage applicable to different building sizes and rate structures will be required for this purpose. A comprehensive optimization method from utility rate structure perspective for such systems will be advantageous in this cause.

### B. Related Works

Many research works have been done on the optimization of the building or other loads using distributed energy resources (DERs) such as solar generation, battery energy storage systems (BESS), and electric vehicles (EV). Minimizing the energy cost of the building-integrated microgrid is the main objective in most of the literature. Thermal modeling of a building and interactive load management [2] and building to the grid scheme are shown in [3], with a view to minimizing the time of use

(TOU) energy cost. The compromise [4] and optimality [5], between user comfort and energy optimization, are executed in terms of energy prices to provide a universal model-based anticipative building energy management system. Plug-in electric vehicles are deployed, and two types of rates are used to provide the optimal strategies [6-7]. Mixed Integer Linear Programming (MILP) approaches are taken to optimize home energy management system, demand response with different electricity prices [8-10]. Though their economic cost function includes complex costs such as fuel cost, the start-up cost of generation units, operation and maintenance cost, etc., the cost paid to the utility is only TOU energy cost. In reality, industrial size building loads almost always include a demand cost which charges for the peak demand that occurs within the billing month.

Demand charges are also analyzed on several occasions. There are two main objectives that are prevalent in demand charge-based research papers. One is focused on optimizing the cost while considering the demand charges and the other is modifying the tariffs to benefit both the consumers and the utilities. In [11], both the demand and energy charges are combined but the formulation is not detailed and easily applicable for calculating the TOU demand part. Critical Peak Pricing (CPP) has also been introduced by the utilities to provide monetary rewards to the consumers for reducing their usage during grid congestion. The impacts of V2G on commercial buildings are investigated during CPP events [12]. Only the effect of a high CPP energy rate is presented during the CPP event hours while the benefits derived from lower demand charge at non-CPP event hours are not shown. An open-source tool [13] is developed to analyze the tariff impacts on the consumers. Effects of the tariffs on the optimal BESS sizing are also shown for a residential load coupled with a solar generation [14]. Their optimization is done incorporating the demand charge through residential loads, but the residential loads usually don't have it. Optimal BESS size and cycle scheduling are sought alongside the combination of both TOU energy and demand charge for residential load in [15]. All these works while adding the effect of demand charge show the analysis only for a single demand charge. Whereas the TOU or time-related demand charges are not uncommon in industrial loads, that charges based on the peak demand occurring at some specific period of the day. In [16-17], new rates are defined for the maximum demand tariff that can satisfy most of the consumers and the utilities. Incentive pricing to maximize the benefits from participating in energy communities is also offered in [18]. A fair pricing mechanism based on forecasted power demand is used to reduce bills for low energy consumers [19]. Economic feasibility is analyzed for user-side energy management based on different energy prices [20-21]. Adjustment of tariff levels is also explored to increase the penetration of DER [22]. A dynamic feed-in tariff is also modeled to replace the flat rate tariffs for photovoltaic integration based on different factors [23]. A segmented energy tariff is designed to flatten the load demand profile [24]. The way energy management system can impact the electricity retail portfolios is analyzed in [25]. Such game theory based complex fair game strategies and a single rate structure based modified electricity rate proposal have failed to capture the necessity of a widespread and accommodating rate structure implementation.

*C. Summary of Contributions*

From the discussions above we can see that while there are ample works on the electricity cost optimization of loads with renewable and BESS, they fail to address the fact that there are numerous and diverse types of utility electricity rates that require distinct optimization strategy based on their type. Much attention is needed on this topic since an optimization strategy for one type of rate structure will not be suitable for the other type and may end up incurring a non-optimal high cost. One may ask several questions in terms of the rate structure: 1) What will be the best rate structure for a particular building load? 2) What will be the best rate structure if the building is equipped with battery energy storage and solar? 3) How much savings are possible for each of the rate structure available with or without renewables and BESS? 4) What will be the projected benefits in future? 5)What charges will impact the savings provided by the optimization? This paper tries to answer all these questions and bridges the gap in literature. Though there are works that can be found in different numbers on different rate structure-based optimizations, to the best of the authors' knowledge no work has been done yet on a comprehensive optimization model that addresses the issue of this complexity and diversity of these utility rates and aims to solve it.

The main contributions of this paper can be summarized as follows:

1. Proposing a comprehensive framework for building load optimization using renewables and energy storage, ubiquitous to any type of rate structure.

2. Detailed modeling of solar fed stationary battery energy storage and system power balance, in compliance with system constraints and configuration.

3. Formulating a novel algorithm to carry out the optimization and generate the desired results systematically.

4. Investigating effects of these various rate structures applied to different building load types and sizes and resulting in the best rate structure recommendation for a building user.

5. Performing a sensitivity analysis to show the relationship between the rate structure and the degree of renewable adoption and BESS deployment along with the examination of the future benefits of the optimization method by doing a projection of rates for the coming years using historical rate structure data.

All these analyses help customers to decide among different rate structure options as well as with resource planning of their renewable generation and energy storage size. This work may also provide a direction to the utility operators to adjust their electricity rates and help reduce the overall grid congestion.

*D. Paper Organization*

This paper is organized as follows: section II gives an overview of the system considered and shows detailed modeling of BESS and system power balance, section III classifies and describes the types of utility rate structures and formulates cost functions for each type, section IV presents the formulation of optimization problem, section V develops the algorithm for carrying out optimization and generating results, section VI discusses the results obtained from the analyses, section VII draws the conclusion and talks about the scope of future works.

## II. SYSTEM CONFIGURATION AND MODELING

### A. Overview of the System

The system we are considering here is a building equipped with a renewable generation such as solar and battery energy storage system (BESS). All the buildings considered in this paper are commercial buildings and have electrical load sizes equivalent to industrial size loads, therefore falling under industrial rate structure.

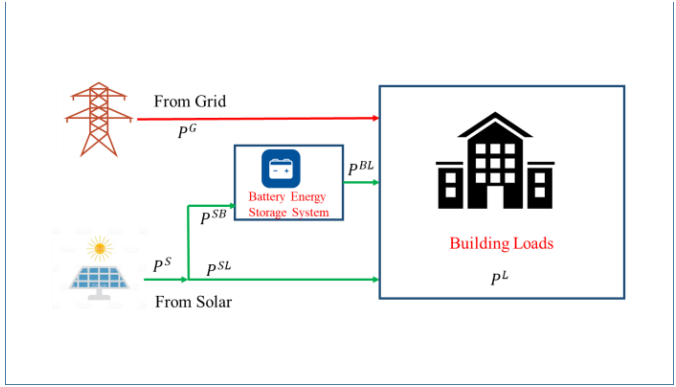

Fig. 1. System Configuration Block Diagram

Figure 1 shows the block diagram of the system configuration. The building load receives some of the power from the utility grid. The power generated from the solar inverter is branched out into two portions. One branch delivers power directly to the building load. The other branch delivers the power into the BESS. Then the BESS delivers power to the building load as required. This way higher efficiency can be achieved by reducing power conversion-associated losses. To ensure a higher percentage of renewable generation within the power mix, the BESS does not store any energy from the grid.

### B. Battery Energy Storage System (BESS) Modeling

The BESS is modeled by taking many real-world operational constraints into considerations. Let $E_t^B$ and $P_t^B$ the energy stored and power that is charged/discharged at time step t, respectively. The battery must maintain its stored charge within a certain limit meaning that the stored energy should be kept within a certain limit at every time step. These limits are imposed in order to increase the lifetime and heath of the BESS. Then,

$$E^{Bmin} \leq E_t^B \leq E^{Bmax} \quad (1)$$

Where, $E^{Bmin}$ and $E^{Bmax}$ are the minimum and maximum limit for stored energy in the battery.

The stored energy at each time step can be calculated with the following equation:

$$E_{t+1}^B = (1-\gamma)E_t^B + P_t^B . \Delta t \quad (2)$$

Where Δt is the duration of each time step and γ is the self-discharge rate. Usually, there is not a linear relationship between the state of charge (SOC) and the stored energy but for a small time step duration, we can assume a linear relationship. The typical value for γ for Li-ion battery is 1-2% over a month [26].

So, if Δt becomes small like 15 minute or 1-minute then γ≈0 and we can rewrite the equation as:

$$E_{t+1}^B = E_t^B + P_t^B . \Delta t \quad (3)$$

The charging/discharging power can be written as:

$$P_t^B = P_t^{B+} - P_t^{B-} \quad (4)$$

Where $P_t^{B+}$ and $P_t^{B-}$ are the charging and discharging power, respectively. These two power quantities should also be within some limitations depending on the size, application, specification, and manufacturer of the BESS inverter. It can be modeled with the following inequality conditions:

$$0 \leq P_t^{B+} \leq P^{B+max} \quad (5)$$

$$0 \leq P_t^{B-} \leq P^{B-max} \quad (6)$$

Where, $P^{B+max}$ and $P^{B-max}$ are the maximum limit for charging and discharging power respectively. But charging and discharging cannot happen simultaneously. This condition can be modeled by the following equation:

$$P_t^{B+} P_t^{B-} = 0 \quad (7)$$

### C. Power Balace Equations

As mentioned in subsection A, the building load is served by drawing power from the grid, a part of solar power and the power discharged from the BESS. The rest of the solar power is used to charge the battery. Let $P^G, P^S$ and $P^L$ be the power drawn from the grid, power generated by the inverter output of the solar photovoltaic (PV) system, and the building load, respectively. $P^{SB}$ and $P^{SL}$ are the portion of solar power generation by the solar inverter that is fed to the BESS and building load, respectively. $P^{BL}$ is the power that is fed to the load by discharging the battery. Let $\eta^+$ and $\eta^-$ be the charging and discharging efficiency of the BESS, respectively. So, for each time step t, the system should obey the following power balance equations:

$$P_t^{B+} = \eta^+ P_t^{SB} \quad (8)$$

$$P_t^S = P_t^{SB} + P_t^{SL} \quad (9)$$

$$P_t^L = P_t^{SL} + P_t^{BL} + P_t^G \quad (10)$$

$$P_t^{BL} = \eta^- P_t^{B-} \quad (11)$$

## III. FORMULATION OF THE COST FUNCTIONS

In this section, the cost functions or the objective functions of the optimization problems are formulated for different utility rate structures. The following subsections show detailed formulations for six types of rate structures. These are:
  A. TOU energy rate with a monthly peak demand charge,
  B. TOU energy rate with TOU or Time Related (TR) demand charge,
  C. TOU energy rate with both monthly peak demand and TR demand charge,
  D. CPP rate,
  E. Flat energy rate with a monthly peak demand charge and
  F. Only TOU energy charge.

Though the utility rates can be versatile, we can cover almost all industrial utility rates with these six types. Some jargons used in the rates may vary by utility company and location, but the main idea remains the same. This paper uses the traditional technical terms such as On-Peak, Mid-Peak, and Off-Peak times, etc. While formulating the cost functions, careful attention is paid to derive them in a way so that they become convex for the convenience of solving them with the available optimization tools and take less time to solve while ensuring the optimality.

*A. Time of Use Energy (TOU) Rate with a Monthly Peak Demand Charge*

In this utility rate the energy charge is decided based on the period of the day and how much stressed is the grid at that period. Typically, they are separated into three time periods: On-Peak, Mid-Peak, and Off-Peak where the value of the energy charge is from highest to lowest, respectively. How much energy is consumed on each period in the whole month is measured and the customer is billed based on that. For the demand charge part, a high value is charged for the maximum demand that happens within the month for a short or in the billing cycle. The utilities measure the moving average demand of the building for a certain duration, usually 15 minutes, then multiply that high value with the maximum 15-minute demand in a billing month to get the demand cost. Then both are added to get the total charge for monthly usage.

Let $\alpha_t$ be the energy charge at time step t and $\beta$ be the monthly peak demand charge. $\alpha_t$ is equal to the On-Peak energy charge when t is within the On-Peak time period defined by the utility company. Similarly, it will take the value of the Mid-Peak and Off-Peak energy charge when t is within the Mid-Peak and Off-Peak periods respectively. Then we can write the cost function or the objective function as:

$$f(P_t^G) = \Delta t \sum_{t=0}^{T-1} \alpha_t P_t^G + \beta \max(P_t^G) \quad (12)$$

Where T is the total number of timesteps considered. We can write the set of all $\alpha_t$ and $P_t^G$ as a vectors $\alpha$ and $P^G$ such as,

$$\boldsymbol{\alpha} = \begin{bmatrix} \alpha_0 \\ \alpha_1 \\ . \\ . \\ \alpha_{T-1} \end{bmatrix} \text{ and } \boldsymbol{P^G} = \begin{bmatrix} P_0^G \\ P_1^G \\ . \\ . \\ P_{T-1}^G \end{bmatrix} \quad (13)$$

Then we can write the cost function as:

$$f(\boldsymbol{P^G}) = \Delta t\, \boldsymbol{\alpha}^T \boldsymbol{P^G} + \beta \max(\boldsymbol{P^G}) \quad (14)$$

*B. TOU Energy Rate with TOU or Time Related (TR) Demand Charge*

Just like the energy charge $\alpha_t$ in the previous subsection, the demand charge $\beta$ is time-dependent similarly in this case. Instead of the maximum 15-minute demand in the whole billing period, the customer is billed for the maximum demand of the three peak periods. The main complexity here is that we can't just pick the largest power value from the vector $\boldsymbol{P^G}$ for this optimization and try to minimize that like the first method. Rather we have to pick the maximum value from each of the time periods and try to minimize each of them in a way so that the total cost becomes minimum. To resolve it, this paper proposes to introduce three diagonal matrices $\boldsymbol{\beta^{On}}$, $\boldsymbol{\beta^{Mid}}$ and $\boldsymbol{\beta^{Off}}$ to represent each of the periods.

$$\boldsymbol{\beta^{On}} = \begin{bmatrix} \beta_0^{On} & \cdots & 0 \\ \vdots & \ddots & \vdots \\ 0 & \cdots & \beta_{T-1}^{On} \end{bmatrix}, \boldsymbol{\beta^{Mid}} = \begin{bmatrix} \beta_0^{Mid} & \cdots & 0 \\ \vdots & \ddots & \vdots \\ 0 & \cdots & \beta_{T-1}^{Mid} \end{bmatrix} \text{ and } \boldsymbol{\beta^{Off}} = \begin{bmatrix} \beta_0^{Off} & \cdots & 0 \\ \vdots & \ddots & \vdots \\ 0 & \cdots & \beta_{T-1}^{Off} \end{bmatrix} \quad (15)$$

Where the diagonal elements $\beta_t^{On}$ = On-Peak charge for t=On-Peak times and 0 for t=other times. Similarly, $\beta_t^{Mid}$ = Mid-Peak charge for t=Mid-Peak times and 0 for t=other times and $\beta_t^{Off}$ = Off-Peak charge for t=Off-Peak times and 0 for t=other times. All the off-diagonal elements are zero. Now, we can write the cost function as:

$$f(\boldsymbol{P^G}) = \Delta t\, \boldsymbol{\alpha}^T \boldsymbol{P^G} + \max(\boldsymbol{\beta^{On}} \boldsymbol{P^G}) + \max(\boldsymbol{\beta^{Mid}} \boldsymbol{P^G})$$
$$+ \max(\boldsymbol{\beta^{Off}} \boldsymbol{P^G}) \quad (16)$$

Sometimes in the winter season of the utility rates, there is no On-Peak demand and in some other rates, there is a Super Off-Peak rate instead, which is a lower charge than the Off-Peak rate. In some cases, the Off-Peak demand charge is absent in summer rates. In those cases, we can similarly formulate the cost function, by making some adjustments to $\boldsymbol{\alpha}$ and $\boldsymbol{\beta}$ as required.

*C. TOU Energy Rate with both Monthly Peak Demand and TR Demand Charge*

This rate is a combination of the rates described in the previous two subsections. It has both a TR demand charge for the maximum demand in the peak periods and a monthly peak demand charge for the maximum peak happening within the billing month. We can write the cost function for this rate as:

$$f(\boldsymbol{P^G}) = \Delta t\, \boldsymbol{\alpha}^T \boldsymbol{P^G} + \max(\boldsymbol{\beta^{On}} \boldsymbol{P^G}) + \max(\boldsymbol{\beta^{Mid}} \boldsymbol{P^G})$$
$$+ \max(\boldsymbol{\beta^{Off}} \boldsymbol{P^G}) + \beta \max(\boldsymbol{P^G}) \quad (17)$$

Where the notations have the same meaning as shown in the previous two subsections. Though in most cases the Off-Peak demand charge is absent in rate. By simply removing the Off-Peak demand charge part from equation 17 we can apply it to those rates.

*D. Critical Peak Pricing (CPP) Rate*

Critical Peak Pricing or CPP rates offer lower demand rates in non-CPP event days in exchange for very high energy rates in CPP event days. CPP events are usually called when electricity demand peaks due to extreme or unusual temperature conditions. They are usually the hottest summer days, occurring 12 days within the summer season as defined by the utility company. The CPP event hours comprise the evening hours and early part of the night namely 4 PM to 9 PM. The customers are generally notified a day before the CPP event that the next day is announced as CPP day. A non-CPP day will have the energy charge vector $\boldsymbol{\alpha_{non-CPP}} = \boldsymbol{\alpha}$ just like the other rates. For a CPP day, we modify it as $\boldsymbol{\alpha_{CPP}}$ where $\alpha_t$ is equal to the CPP energy charge when t falls within the CPP event hours. The demand charge matrices for the non-CPP days $\boldsymbol{\beta}^{On}_{non-CPP}, \boldsymbol{\beta}^{Mid}_{non-CPP}$ and $\boldsymbol{\beta}^{Off}_{non-CPP}$ need to be modified. We can do that by replacing the t th diagonal element in the matrix with the actual charge minus the discount charge offered by the utility company when t falls under the CPP event hours. The same matrices for CPP days $\boldsymbol{\beta}^{On}_{CPP}, \boldsymbol{\beta}^{Mid}_{CPP}$ and $\boldsymbol{\beta}^{Off}_{CPP}$ are the same as the other TR demand matrices shown in the previous subsections. So, we get two cost functions:

$$f(\boldsymbol{P^G}) = \Delta t\, \boldsymbol{\alpha}_{non-CPP}^T \boldsymbol{P^G} + \max(\boldsymbol{\beta}^{On}_{non-CPP}\boldsymbol{P^G}) + \max(\boldsymbol{\beta}^{Mid}_{non-CPP}\boldsymbol{P^G}) + \max(\boldsymbol{\beta}^{Off}_{non-CPP}\boldsymbol{P^G}) + \beta\max(\boldsymbol{P^G}) \quad (18)$$

and

$$f(\boldsymbol{P^G}) = \Delta t\, \boldsymbol{\alpha}_{CPP}^T \boldsymbol{P^G} + \max(\boldsymbol{\beta}^{On}_{CPP}\boldsymbol{P^G}) + \max(\boldsymbol{\beta}^{Mid}_{CPP}\boldsymbol{P^G}) + \max(\boldsymbol{\beta}^{Off}_{CPP}\boldsymbol{P^G}) + \beta\max(\boldsymbol{P^G}) \quad (19)$$

Where equations 18 and 19 represent the cost functions for non-CPP days and CPP days respectively. Note that, the discounted rates at non-CPP days are also provided during the same hours as the CPP event hours at CPP days.

*E. Flat Energy (TOU) Rate with a Monthly Peak Demand Charge*

For a flat energy rate with a monthly peak demand charge the variable $\alpha_t$ becomes a constant independent of time and we can replace the energy charge vector $\boldsymbol{\alpha}$ with a single scalar value $\alpha$. The cost function then becomes:

$$f(\boldsymbol{P^G}) = \Delta t.\alpha\,.\boldsymbol{1}^T \boldsymbol{P^G} + \beta\max(\boldsymbol{P^G}) \quad (20)$$

Where **1** denotes a vector of size T with all elements as 1. The other notations have the same meaning from previous subsections.

*F. TOU Energy Rate Only*

For only the TOU energy rate without any demand charge, we make the value of monthly peak demand charge $\beta=0$. Then the cost function can be written as:

$$f(\boldsymbol{P^G}) = \Delta t\, \boldsymbol{\alpha}^T \boldsymbol{P^G} \quad (21)$$

IV. OPTIMIZATION PROBLEM FORMULATION

Using the cost functions formulated in section III as the objective function and the BESS modeling equations from subsection B of section II as constraints, we can now derive the optimization problem for each of the rate structures described above.

But equation 7 in the model is a nonlinear equality condition which makes the problem nonconvex and difficult to solve. We can resolve it by introducing a binary variable $\delta_t \in \{0,1\}$ to the above mentioned power inequality constraints. So, we reformulate them as:

$$0 \leq P_t^{B+} \leq \delta_t\, P^{B+max} \quad (22)$$

$$0 \leq P_t^{B-} \leq (1-\delta_t)\, P^{B-max} \quad (23)$$

Though the nonlinearity problem is dealt with still the equations remain nonconvex as the binary variable introduces a separate nonconvexity. But they can be made convex by doing a convex relaxation on the binary variable and write it as:

$$0 \leq \delta_t \leq 1 \quad (24)$$

We can now write the optimization problem as shown below:

$$\min_{\boldsymbol{P^G}} f(\boldsymbol{P^G})$$

subject to:

$$E_{t+1}^B = E_t^B + P_t^B.\Delta t \quad (3)$$

$$E^{Bmin} \leq E_t^B \leq E^{Bmax} \quad (1)$$

$$P_t^B = P_t^{B+} - P_t^{B-} \quad (4)$$

$$0 \leq P_t^{B+} \leq \delta_t\, P^{B+max} \quad (22)$$

$$0 \leq P_t^{B-} \leq (1-\delta_t)\, P^{B-max} \quad (23)$$

$$0 \leq \delta_t \leq 1 \quad (24)$$

$$P_t^{B+} = \eta^+ P_t^{SB} \quad (8)$$

$$P_t^S = P_t^{SB} + P_t^{SL} \quad (9)$$

$$P_t^L = P_t^{SL} + P_t^{BL} + P_t^G \quad (10)$$

$$P_t^{BL} = \eta^- P_t^{B-} \quad (11)$$

By applying the appropriate cost function and parameter values, we can use this optimization model to obtain the optimal operation of any BESS, for any type of industrial rate structure, given a building load and a solar generation profile. Note that, the constraints in the optimization problem are not in their numerical order. This is done to maintain continuity for readers' convenience.

V. ALGORITHM

For our proposed method we do the optimization on a day-by-day basis instead of doing a monthly optimization. We do this for a couple of reasons. Firstly, the day ahead prediction of load and solar can be produced more accurately and easily for a shorter time resolution such as 15 minutes. For this sort of time resolution month ahead prediction will contain more inaccuracy. By doing a day ahead prediction we can also easily determine the operation of the BESS for the next day. Secondly, doing a monthly optimization with this 15-minute data resolution would be computationally exhaustive. Without a very high processing power and large data storage device, the optimization will require much more computational time to solve.

Since we are doing a daily optimization, the total number of time steps considered T will be 96 in the case of 15-minute resolution. The vectors and matrices introduced in section III will have corresponding sizes. We can write the daily building load profile and solar generation profile in vectorized form as:

$$\boldsymbol{P^L} = \begin{bmatrix} P_0^L \\ P_1^L \\ . \\ . \\ P_{T-1}^L \end{bmatrix} \text{ and } \boldsymbol{P^S} = \begin{bmatrix} P_0^S \\ P_1^S \\ . \\ . \\ P_{T-1}^S \end{bmatrix} \quad (25)$$

In this section, we will refer to the daily energy charge and demand charges as $\alpha$ and $\beta$ in general. In order to calculate the actual monthly usage bill, we will need to get new vectors and matrices that will accommodate the monthly calculation. If the number of total days in the monthly billing cycle is D, then we can create the new matrices and vectors of size D×T for the monthly bill calculation. The energy charge vector for the month $\alpha_{month}$ will then become:

$$\alpha_{month} = [\ \alpha^T\ \alpha^T\ \ldots\ldots\ \alpha^T]^T \qquad (26)$$

Where $\alpha$ is repeated D times. We can get the On-Peak TR demand charge matrix for the month $\beta_{month}^{On}$ by:

$$\beta_{month}^{On} = \begin{bmatrix} \beta^{On} & \cdots & 0 \\ \vdots & \ddots & \vdots \\ 0 & \cdots & \beta^{On} \end{bmatrix} \qquad (27)$$

Where $\beta^{On}$ is repeated on the diagonal position D times and $0$ is a square matrix of size T with all elements as zeros. Similarly, we can find the other TR demand matrices $\beta_{month}^{Mid}$ and $\beta_{month}^{Off}$, which represent the Mid-Peak and Off-Peak TR demand charge matrix respectively.

If we are dealing with a CPP rate structure in the summer season, then equations 26 and 27 will be a little different. All the alphas will be replaced by $\alpha_{non-CPP}$ except for position d in the $\alpha_{month}$ vector if day number d is a CPP event day. In that case, it will be replaced by $\alpha_{CPP}$. Similarly, all the betas will be replaced by $\beta_{non-CPP}^{On}, \beta_{non-CPP}^{Mid}$ and $\beta_{non-CPP}^{Off}$ as in their respective monthly matrices except for d th position where they will be replaced by $\beta_{CPP}^{On}, \beta_{CPP}^{Mid}$ and $\beta_{CPP}^{Off}$, respectively.

We can then calculate the monthly usage bills by using the applicable cost function just by using the appropriate power vector of the right size and parameters $\alpha_{month}$ and $\beta_{month}$.

$$Monthly\ Usage\ Bill = f(X; \alpha_{month}, \beta_{month}) \qquad (28)$$

Where **X** is the appropriate power vector of size D×T. We can now calculate the monthly usage bill for a building load alone, a building net load with solar or net with solar and BESS. To better understand the effects of the optimization, we define two types of savings, Savings 1 and Savings 2. Savings 1 shows the savings if only the solar is added. Savings 2 shows the effect of BESS optimization that is the additional savings we made by BESS optimization when solar is already present in the system. We can write them as:

$$Savings\ 1 = f(P_{month}^L; \alpha_{month}, \beta_{month}) - f(P_{month}^L - P_{month}^S; \alpha_{month}, \beta_{month}) \qquad (29)$$

$$Savings\ 2 = f(P_{month}^L - P_{month}^S; \alpha_{month}, \beta_{month}) - f(P_{month}^G; \alpha_{month}, \beta_{month}) \qquad (30)$$

To run the optimization, we have to get the other parameters $E^{Bmax}, E^{Bmin}, P^{B+max}, P^{B-max}, \eta^+$ and $\eta^-$. We would also need the initial stored energy $E_{init}^B$ of the BESS. Variables $P^{SL}$, $P^{SB}, P^{B+}$ and $P^{B-}$ need to be declared which are similar to vectors shown in equation 25.

Now that we have described all the necessary variables and parameters, the algorithm for the optimization process can be written as:

| Algorithm |
|---|
| 1: Select the type of Rate Structure |
| 2: Select Season |
| 3: Get parameters $E^{Bmax}, E^{Bmin}, E_{init}^B, P^{B+max}, P^{B-max}, \eta^+$ and $\eta^-$. |
| 4: **if** (Rate type=CPP rate) **and** (Season=Summer) **then**: |
| 5:     Get $\alpha_{CPP}, \beta_{CPP}, \alpha_{non-CPP}$, and $\beta_{non-CPP}$ and the CPP cost function $f(.)$ |
| 6: **else** |
| 7:     Get $\alpha, \beta$ and the cost function $f(.)$ |
| 8: **endif** |
| 9: Declare empty vectors $P_{month}^G = [\ ], P_{month}^L = [\ ], P_{month}^S = [\ ]$ |
| 10: **for** each day d=0,1,……., D-1 **do**: |
| 11:     Get $P^L, P^S$ for day d |
| 12:     Declare the other variables $P^{SL}, P^{SB}, P^{B+}$ and $P^{B-}$ |
| 13:     Initialize $E_0^B = E_{init}^B$ |
| 14:     Run Optimization |
| 15:     Run the optimum BESS operation according to $P^{B+}$ and $P^{B-}$ |
| 16:     Update $P_{month}^G = \begin{bmatrix} P_{month}^G \\ P^G \end{bmatrix}$ |
| 17:     Update $P_{month}^L = \begin{bmatrix} P_{month}^L \\ P^L \end{bmatrix}$ and $P_{month}^S = \begin{bmatrix} P_{month}^S \\ P^S \end{bmatrix}$ |
| 18:     Update $E_{init}^B = E_{T-1}^B$ |
| 19: **endfor** |
| 20: Calculate the unoptimized monthly cost $f(P_{month}^L; \alpha_{month}, \beta_{month})$ |
| 21: Calculate the unoptimized monthly cost if solar is added $f(P_{month}^L - P_{month}^S; \alpha_{month}, \beta_{month})$ |
| 22: Calculate the optimized monthly cost with solar and BESS $f(P_{month}^G; \alpha_{month}, \beta_{month})$ |
| 23: Calculate Savings 1 and Savings 2 |

## VI. RESULTS AND DISCUSSIONS

Data for four building loads are collected for simulation of the optimization model. All these buildings are commercial buildings and fall under different types of industrial rate structures. These buildings differ in electrical usage and functionality. Among them three of the buildings have actual onsite solar generation and two of them have actual BESS integrated. For simulation, all the building loads that are used are actual building load data. Except for CPP, the rates that are used are actual utility rates these buildings are billed for. For two of these buildings, 15-minute data were readily available. For the other two, available 1-minute data were averaged for every 15 minutes to get the 15-minute data. For the buildings which did not have solar or BESS, National Renewable Energy Laboratory or NREL's REopt Lite was used to find the ideal solar or BESS size for the corresponding building [27]. The actual building load data and if available, actual solar data were given as inputs and simulated to get the output ideal size. As the size of solar or BESS from the original simulated output value may not be easily available in the real world, reasonably close commercially available sizes were selected. Then using that simulated size NREL's System Advisor Model (SAM) was used to generate the solar profile for the site where a solar generation was absent [28]. Table I summarizes the characteristics of the buildings that were used for the optimization which are: average daily energy usage, load factor (average load divided by maximum load in a period), rate structure, solar and BESS size. The rate structure types mentioned in the table are referred to in section III.

TABLE I. CHARACTERISTICS OF THE BUILDINGS USED IN OPTIMIZATION

| Characteristics | Building 1 | Building 2 | Building 3 | Building 4 |
|---|---|---|---|---|
| Average Daily Usage (kWh) | 11,223 | 1,660 | 13,451 | 805 |
| Load Factor | 0.283 | 0.341 | 0.811 | 0.411 |
| If Real Solar | Yes | Yes | No | Yes |
| If Real BESS | No | Yes | No | Yes |
| Solar Size (Real) | 800 kW | 220 kW | - | 180 kW |
| BESS Size (Real) | - | 100 kW/ 500 kWh | - | 100 kW/ 500 kWh |
| Solar Size (Simulated) | - | - | 650 kW | - |
| BESS Size (Simulated) | 150 kW/ 500 kWh | - | 120 kW/ 320 kWh | - |
| Rate Structure | Type A | Type B | Type C | Type E |

Using the data and parameters collected and simulated the optimization for these four buildings were done for their corresponding rate structures. All the data were used are from the summer months and the rates used were summer season rates. For all these rates the daily On-Peak hours were from 12 PM to 6 PM, Mid-Peak hours were from 8 AM to 12 PM and from 6 PM to 11 PM and the remaining hours are all Off-Peak Hours. At the start of each simulation, it is assumed that all the BESS have their initially stored energy $E_{init}^B$ at 50% of the total capacity. The limits for maximum and minimum stored energy for each BESS $E^{Bmax}$ and $E^{Bmin}$ is assumed to be 90% and 20% of their total capacity, respectively. Figure 2 shows the results of the optimization. For each of the simulations, the building load without optimization and the load optimized with solar and BESS for a month are shown. For the simulations in this work Matlab based convex optimization toolbox CVX have been used [29].

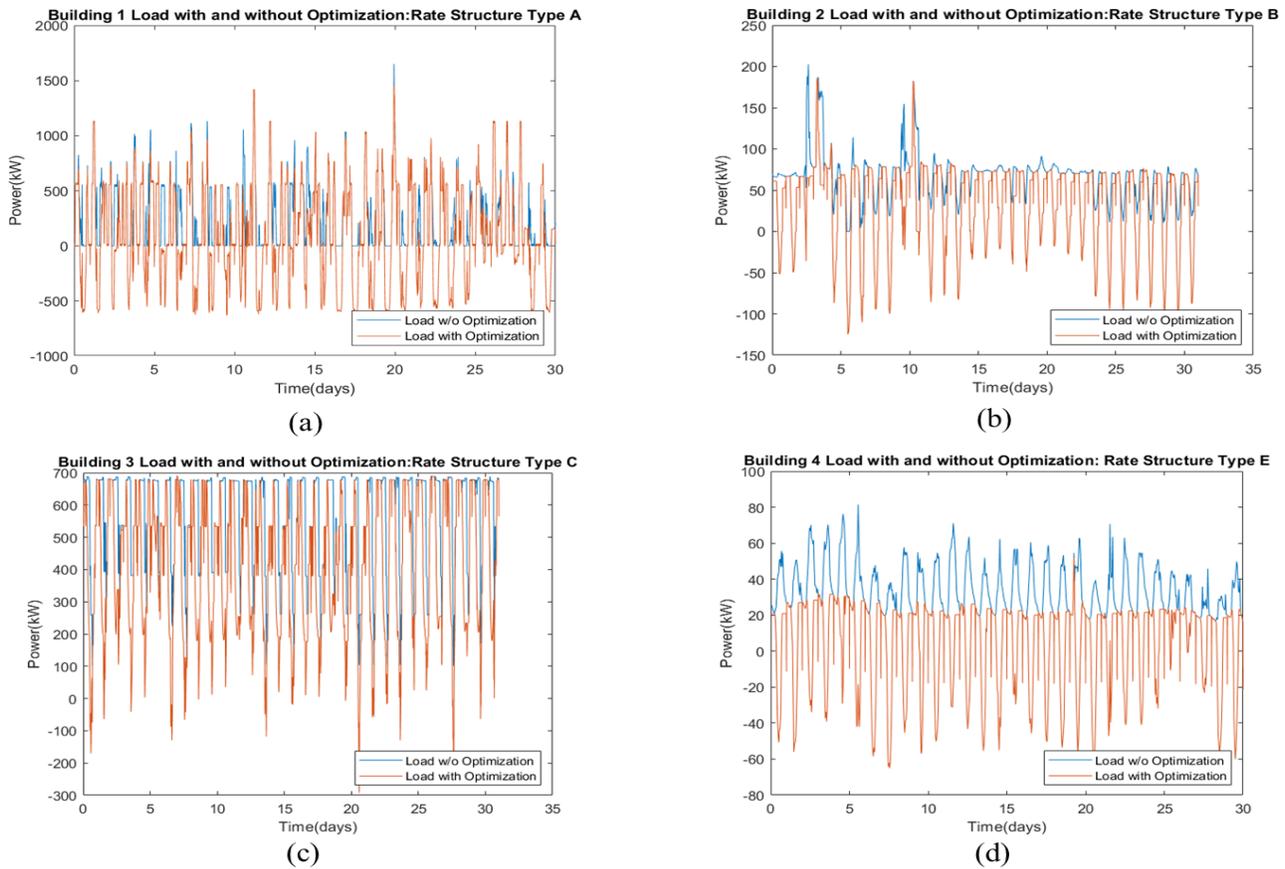

Fig. 2. Results from Optimization of (a) Building 1 with Rate Structure Type A, (b) Building 2 with Rate Structure Type B, (c) Building 3 with Rate Structure Type C, (d) Building 4 with Rate Structure Type E

For rate type D or CPP, the utilities normally give customers with some rate structures an option that they can either move to CPP or can stay under the existing rate structure. It may be opt-in or opt-out. The customers decide what option they want. In this paper Building 1 is chosen for CPP rate simulation and the rates for the CPP option for that rate structure are applied here. This rate has four months in the summer season (June to September) and 12 CPP event days in the summer season in total. So, on average each month will have three CPP days. The three highest demand days (day numbers 9, 12, and 20) are selected as CPP event days in this simulation. The CPP event hours take place from 4 PM to 9 PM. Figure 3 shows the results of the simulation where loads with and without optimization similar to figure 2 for other buildings are shown.

TABLE II.    SUMMARY OF OPTIMIZATION RESULTS

| Optimization Parameters and Results | Type A | Type B | Type C | Type D | Type E |
|---|---|---|---|---|---|
| Building | Building 1 | Building 2 | Building 3 | Building 1 | Building 4 |
| Energy Charge ($/kWh) | On-Peak: 0.3397<br>Mid-Peak: 0.13837<br>Off-Peak: 0.07637 | On-Peak: 0.35987<br>Mid-Peak: 0.1007<br>Off-Peak: 0.03545 | On-Peak: 0.10258<br>Mid-Peak: 0.07566<br>Off-Peak: 0.05727 | On-Peak: 0.07817<br>Mid-Peak: 0.07422<br>Off-Peak: 0.0724<br>CPP: 0.4 | Flat rate: 0.0139 |
| Demand Charge ($/kW) | Monthly Peak: 11.87 | On-Peak: 7.06<br>Mid-Peak: 3.13<br>Off-Peak: 1.53 | On-Peak: 21.73<br>Mid-Peak: 4.17<br>Monthly Peak: 19.02 | On-Peak: 16<br>Mid-Peak: 5.16<br>Monthly Peak: 17.52<br>CPP discount: 4.11 | Monthly Peak: 10.58 |
| Unoptimized Cost ($) | 48,825 | 9,278 | 61,704 | 79,714 | 4,218 |
| Unoptimized Cost with Solar ($) | 14,849 | 1,911 | 50,766 | 68,778 | 1,122 |
| Optimized Cost with Solar and BESS ($) | 12,483 | 1,270 | 49,250 | 63,067 | 1,064 |
| Savings 1 (Additional Savings from Solar) ($) | 33,976 | 7,368 | 10,938 | 10,936 | 3,096 |
| Savings 2 (Additional Savings from BESS) ($) | 2,366 | 641 | 1,516 | 5,711 | 58 |

The rate type F, the one where there is no peak demand charge is not simulated in this paper. Though these types of rates are not entirely obsolete and still exist in some of the utilities, the commercial buildings where they are applied are rarely found at present. But if there is building data available that falls under this rate structure, it can be simulated using an optimization model and algorithm from the previous sections.

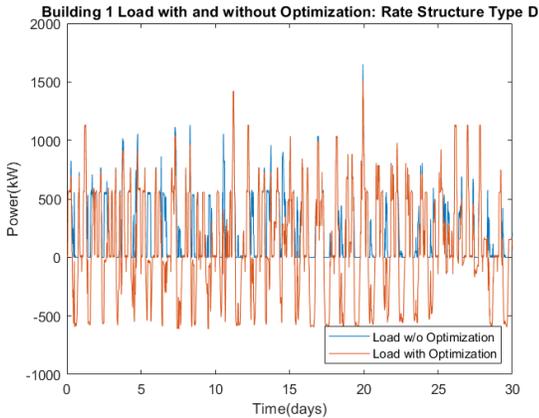

Fig. 3.  Results from Optimization of Building 1 with Rate Structure Type D

Table II summarizes the optimization results from all the simulations. From the table, we can see that building 3 with rate type C has much lower energy charges than building 1 with rate type A. Still building 3 has much higher unoptimized cost than building 1 whereas they have comparable daily kWh usage. This can be attributed to the additional TR demand charges of building 3. But we can see that the savings from adding solar and BESS is higher in building 1 which can be attributed to the fact that building 1 has a lower load factor than building 3 while having higher solar and BESS sizes than the latter. Both rate types A and D were applied on building 1. Both the unoptimized and optimized costs were much greater for rate type D than type A. While adding solar provided more savings in the case of rate type A, adding BESS provided higher savings in case of type D. So, we can see that the CPP option would not have been beneficial to that customer even with solar and optimal BESS operation. Although if it were made mandatory, then adding BESS with solar would serve to be more beneficial. Building 2 has almost double daily kWh usage compared to building 4 while having a somewhat close load factor. The average of the energy costs for three peak periods of building 2 is close to the flat energy cost of building 4, while the sum of demand charges of the former is close to the single demand charge of the latter. They also have the same size solar and BESS. These are reflected in the fact that the unoptimized costs and savings from solar for building 4 are almost twice as building 2. But the savings from adding BESS is almost 11 times in building 4 compared to building 2. It shows that rate type B can be more beneficial by adding BESS compared to rate type E.

Another important issue with the utility rates is that the charges change every year. To understand how these changes impact the savings and how much benefit we will get in the future from this optimization, a sensitivity analysis was done on the building 1 load data using historical values of rate type A. Using the historical energy and demand charge values of the summer season from last 12 years a projection of the next 3 years was made. Figures 4 and 5 show the historical and projected energy charges and monthly peak demand charges, respectively. A second-order polynomial fit was used to execute the projections.

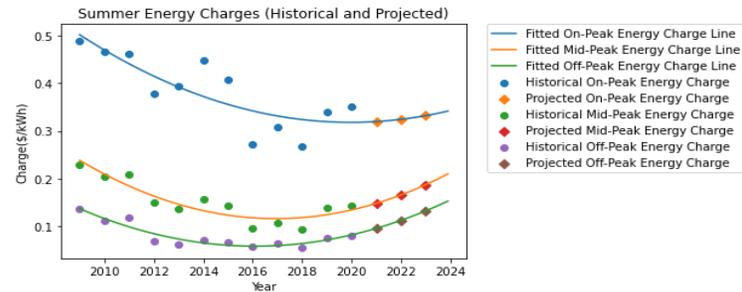

Fig. 4.  Historical and Projected Summer Energy Charges

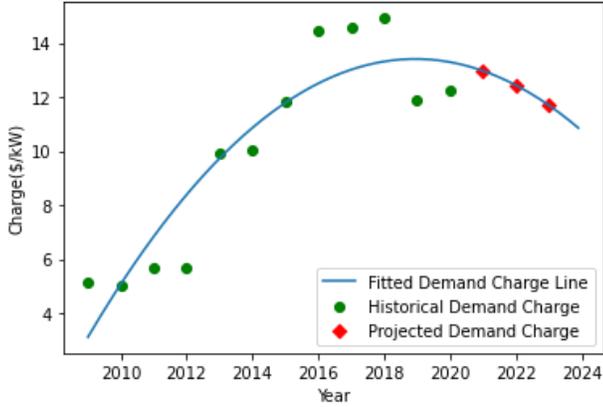

Fig. 5. Historical and Projected Summer Monthly Peak Demand Charges

Using the historical and projected energy and demand charge values the optimization was carried out for all 15 years. Figures 6 and 7 show savings 1 and savings 2 for each of the years. Table III shows the Pearson correlation coefficients of the savings with the energy and demand charges to find out how the change in these charges affect the savings from adding renewable energy or BESS. Positive or negative values of the coefficients denote that the variables are positively or negatively correlated while the absolute value close to zero or one denotes weak or strong correlation, respectively.

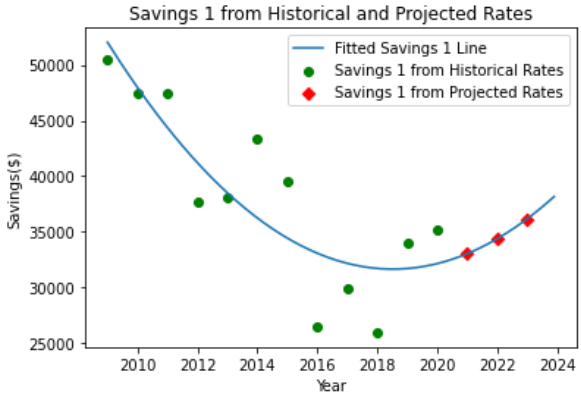

Fig. 6. Historical and Projected Savings 1

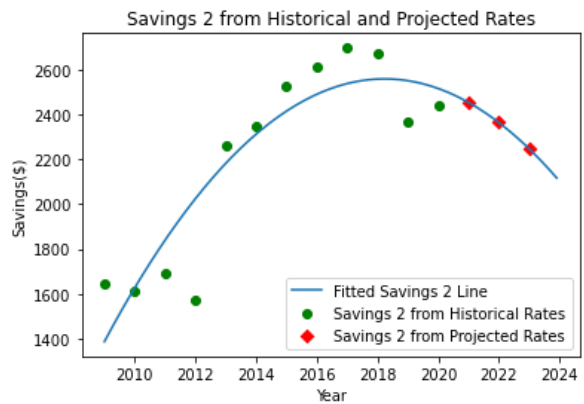

Fig. 7. Historical and Projected Savings 2

TABLE III. PEARSON CORRELATION COEFFICIENTS AMONG THE SAVINGS AND THE UTILITY CHARGES

| Savings | On-Peak Energy Charge | Mid-Peak Energy Charge | Off-Peak Energy Charge | Demand Charge |
|---|---|---|---|---|
| Savings 1 | 0.969 | 0.842 | 0.545 | -0.996 |
| Savings 2 | -0.948 | -0.88 | -0.606 | 0.986 |

As we can see from figures 6 and 7 and the values from table III, savings 1 or savings from adding solar only is dominated by the change in the energy charges, specially the On-Peak charge, while savings 2 or the additional savings from the BESS when solar is already present, is dominated by the changes in monthly peak demand charge.

VII. CONCLUSION AND FUTURE WORK

In this paper, a comprehensive optimization method was developed for commercial buildings equipped with renewable generation and BESS to capture the diversity in utility rate structure types. A BESS model was generated for this purpose. The utility rate structures were categorized into six universal rate structure types that cover almost all the industrial rate structures. Then for each of the types, a cost function was formulated. Using the cost functions and the BESS model an optimization problem was formulated along with an algorithm to apply it on commercial buildings with solar generation and BESS to generate the optimal BESS operation profile. The optimization was done on four buildings using five of the rate structure types that are most common. Results show that TR demands and CPP charges cause higher unoptimized costs. Buildings with lower load factors will gain higher savings from adding solar and BESS. In the case of CPP while only solar may not result in being more beneficial, adding BESS with optimization will provide increased additional benefits. Adding BESS would provide much lower savings in case of flat energy rate than TOU energy rates. A sensitivity analysis of energy and demand charge changes on the optimization was also executed. For that 12 years of historical data were used to do a projection for the upcoming three years. The results revealed that savings from adding solar were dominated by the energy charge changes while the additional savings from optimal BESS operation was dominated by demand charge changes. So, if the energy charge increases in the future adding more solar generation will be beneficial whereas increasing the size of BESS would result in more potential savings if the demand charge is increased. Here, actual building load data and solar data, either actual or simulated, that are used in the simulations all correspond to their authentic timestamps. But when doing a day ahead optimization in real life, we have to use predicted building load and solar generation data. While a good prediction should generate very close results with the original ones, a bad one may cause serious deviations from those. The authors would like to do a prediction on the building loads and solar generations, do optimizations using the predicted data, and investigate how they compare with the original results. Increasing EV adoption and on campus charging infrastructures in workplaces along with vehicle to grid (V2G) capabilities have brought changes in the shape of commercial building loads, subsequently causing utilities to introduce EV rates. Though these EV loads are separately metered, the prospects of coordinated charging with V2G working as a BESS of the workplace buildings under building rate structure, needs to be assessed. More future research can be done on this aspect.